\documentclass[showpacs,amsmath,amssymb,twocolumn,prapplied,reprint,superscriptaddress]{revtex4-2}
\usepackage[dvips]{graphicx} 
\usepackage{amsfonts}
\usepackage{amssymb}
\usepackage{amscd}
\usepackage{amsmath}    
\usepackage{enumerate}
\usepackage{epsfig}
\usepackage{subfigure}
\usepackage{bm}
\usepackage{xcolor}
\usepackage{amsthm}
\usepackage{framed}
\usepackage{multirow}
\usepackage{mathrsfs,amssymb}
\usepackage{latexsym} 
\usepackage{physics}
\usepackage{epstopdf}
\usepackage[skins]{tcolorbox}
\newtcolorbox[auto counter]{tbox}[2][]{%
    enhanced, float=hbt, drop fuzzy shadow southeast,
    colback=white!5!white, colframe=white!50!black,
    width= .97\columnwidth,sharp corners, boxrule=0.8pt,
    title={Table \thetcbcounter: #2}, #1
}
\usepackage[linesnumbered,boxed,ruled,vlined,rightnl]{algorithm2e}
\usepackage{qcircuit}

\begin{document}
\title{A hybrid algorithm for quadratically constrained quadratic optimization problems}

\begin{abstract}
Quadratically Constrained Quadratic Programs (QCQPs) are an important class of optimization problems with diverse real-world applications. In this work, we propose a variational quantum algorithm for general QCQPs. By encoding the variables on the amplitude of a quantum state, the requirement of the qubit number scales logarithmically with the dimension of the variables, which makes our algorithm suitable for current quantum devices. Using the primal-dual interior-point method in classical optimization, we can deal with general quadratic constraints.
Our numerical experiments on typical QCQP problems, including Max-Cut and optimal power flow problems, demonstrate a better performance of our hybrid algorithm over the classical counterparts.
\end{abstract}

\author{Hongyi Zhou}
\email{zhouhongyi@ict.ac.cn}
\affiliation{State Key Lab of Processors, Institute of Computing Technology, Chinese Academy of Sciences, 100190, Beijing, China.}

\author{Sirui Peng}
\affiliation{State Key Lab of Processors, Institute of Computing Technology, Chinese Academy of Sciences, 100190, Beijing, China.}

\author{Qian Li}
\email{liqian.sea@hotmail.com}
\affiliation{Shenzhen lnternational  Center For Industrial  And  Applied  Mathematics, Shenzhen Research Institute of Big Data, Shenzhen, China.}

\author{Xiaoming Sun}
\email{sunxiaoming@ict.ac.cn}
\affiliation{State Key Lab of Processors, Institute of Computing Technology, Chinese Academy of Sciences, 100190, Beijing, China.}

\maketitle
\section{Introduction}

Quadratically constrained quadratic programs (QCQPs) involve optimizing a quadratic objective function subject to quadratic constraints, which are an important class of optimization problems with wide-ranging applicability in modeling and solving various real-world problems across diverse fields. Typical examples include the Maximum Cut (Max-Cut) problems and boolean optimization in computer science, optimal power flow (OPF) problems in electrical engineering, portfolio optimization problems in finance \cite{cui2013convex}, steering direction estimation for radar detection \cite{de2010fractional}, and triangulation in computer vision \cite{aholt2012qcqp}, among others.

The primary challenge in solving QCQPs stems from their non-convex nature. Classical solvers often encounter local minima, hindering their effectiveness in providing globally optimal solutions. As a result, searching for the global optimal solution becomes a severe task, particularly in high-dimensional spaces, where the computational complexity may scale exponentially with the dimension of the variables. There are also various approximation algorithms proposed to circumvent this problem by transforming the non-convex optimizations into convex ones, such as the semi-definite relaxation approach \cite{luo2010semidefinite,Boyd2006ConvexO} and the reformulation linearization technique \cite{anstreicher2012convex}. However, all of these approximation algorithms have limitations in certain cases \cite{mehanna2014feasible}. 
In fact, efficient solutions are known for only specific classes of QCQPs \cite{Boyd2006ConvexO}. Solving general QCQPs is NP-hard, leaving a notable gap in terms of computationally efficient solvers. Thus, addressing general QCQPs on a large scale remains a significant and formidable challenge.

Recent advancements in quantum technology have introduced promising avenues for tackling QCQPs. Quantum algorithms have shown superior performance in solving specific problems with significantly reduced computational complexity. Among them, the quantum approximate optimization algorithm (QAOA) \cite{farhi2014quantum} provides a way to deal with a special type of QCQP, the quadratic unconstrained binary optimization (QUBO), where the variables can only take 0 or 1. The binary variables are encoded in the spin information of a physical system. Then the optimization problem is transformed into finding the ground-state energy. The QAOA algorithm demands a number of qubits linearly proportional to the problem's dimension, which poses challenges for solving large-scale QCQPs. Another quantum algorithm for QUBO is based on semi-definite relaxation \cite{brandao2022faster}. This approach involves solving a semi-definite program as an approximation of the original QUBO problem, employing the quantum Gibbs sampling algorithm \cite{brandao2017quantum}. However, implementing this on current quantum devices remains a substantial challenge. Consequently, designing a quantum algorithm for general QCQPs that can be realized on near-term quantum devices remains an open research question.

In this work, we propose a variational quantum algorithm for QCQPs. Compared with previous works, our algorithm extends the applicability to a broader range of QCQPs beyond the scope of QUBO, accommodating general quadratic constraints. Meanwhile, our algorithm requires far fewer qubit resources that can be easily implemented on current quantum devices. Such advantages are based on the following two main ideas.
By encoding the variables in the amplitude of a quantum state, the requirement of the qubit number scales logarithmically with the dimension of the variables.
Using the primal-dual interior-point method in classical optimization, we can deal with the general constraints by constructing the Lagrange function
and optimize primal and dual variables simultaneously.
To validate the effectiveness of our approach, we conduct numerical experiments on typical QCQP problems, including the Max-Cut problem and the Optimal Power Flow problem. Our results demonstrate that our variational quantum algorithm can outperform classical algorithms, providing improved solutions. This research marks a significant step towards harnessing the power of quantum computing for solving QCQPs on a large scale
efficiently and effectively.

\section{Preliminary}
\subsection{Quadratically Constrained Quadratic Program}
We introduce the general forms of both complex-valued and real-valued QCQP. Suppose the number of variables in both cases is $N$. Without loss of generality, the general form of a complex-valued quadratically constrained quadratic program (QCQP) is given by a homogeneous form \cite{mehanna2014feasible},
\begin{equation}\label{eq:qcqp}
\begin{aligned}
 \min_{\boldsymbol{x}\in \mathbb{C}^N} \quad &  \boldsymbol{x}^\dag A_0 \boldsymbol{x} \\
\mathrm{s.t.} \quad &   \boldsymbol{x}^\dag A_i \boldsymbol{x} \leq c_i , \quad i\in\{1,2,\cdots,m\},
\end{aligned}
\end{equation}
where $A_i$ $(i\in \{0,1,2,\dotsc,m\})$ is Hermitian but possibly indefinite. When all $A_i$ are positive semi-definite, the problem Eq.~\eqref{eq:qcqp} is equivalent to a semi-definite programming (SDP) problem that can be efficiently solved.

For real-valued QCQP, we also assume a general homogeneous form \cite{Park2017GeneralHF},
\begin{equation}\label{eq:qcqpreal}
\begin{aligned}
 \min_{\boldsymbol{y}\in \mathbb{R}^N} \quad &  \boldsymbol{y}^T B_0 \boldsymbol{y} \\
\mathrm{s.t.} \quad &   \boldsymbol{y}^T B_i \boldsymbol{y} \leq d_i , \quad i\in\{1,2,\cdots,m\},\\
\end{aligned}
\end{equation}
where $B_i$ $(i\in \{0,1,2,\dotsc,m\})$ are symmetric matrices. Real-valued QCQPs are much more expressive than complex-valued QCQPs. See \cite{Boyd2006ConvexO} for examples that can be modeled as real-valued QCQP. In particular, it is a basic fact that a complex-valued QCQP can be recast as a real-valued one \cite{eltved2021convex}. 

\subsection{Variational Quantum Algorithm}
In a variational algorithm, a parameterized quantum circuit and a classical optimizer are employed. A variational algorithm is essentially an iterative algorithm optimizing the circuit parameters, where the target function, also known as the cost function, and its gradient are computed by performing various measurements on the quantum circuit. In each iteration, the search direction of the circuit parameters is computed on the classical computer and updated in the quantum circuit.
Once the termination condition of the algorithm is satisfied, the variational algorithm outputs an estimate of the solution to the problem. If we denote the circuit parameters as a vector $\boldsymbol{\theta} \in \mathbb{R}^n$, 
then the output state of the quantum circuit and the cost function can be expressed as $\ket{\psi(\boldsymbol{\theta})}$ and $F(\boldsymbol{\theta})$, respectively. The optimization problem of a variational algorithm is given by $\min_{\boldsymbol{\theta}\in \mathbb{R}^n} F(\boldsymbol{\theta})$. A simple example of the cost function can be $F(\boldsymbol{\theta}) = \bra{\psi(\boldsymbol{\theta})}O\ket{\psi(\boldsymbol{\theta})}$ for an observable $O$.

\section{Main Result}

\subsection{Estimating the Quadratic Form by a Quantum Computer}

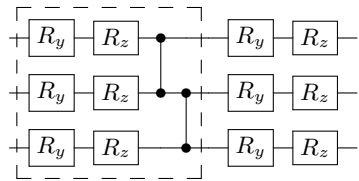
\begin{figure}[!ht]
     \centerline{
        \Qcircuit @C = 0.8 em @R =0.75 em  {
                & \lstick{}& \gate{R_y} & \gate{R_z} 
                &\ctrl{1} & \qw  &\qw  &\gate{R_y}              & \gate{R_z} & \qw 
                \\
               & \lstick{}& \gate{R_y} & \gate{R_z} 
               & \control \qw & \ctrl{1}       &\qw     &\gate{R_y}               & \gate{R_z} & \qw 
               \\ 
            & \lstick{}& \gate{R_y} & \gate{R_z} 
                & \qw &  \control \qw &       \qw &       \gate{R_y}             & \gate{R_z} &\qw 
               \gategroup{1}{3}{3}{6}{1em}{--}
            }
    }
    \caption{A typical implementation of the hardware-efficient ansatz. Each single-qubit rotation gate is parameterized by a single scalar $\theta_i$, which is the $i$-th component of $\boldsymbol{\theta}$. The gates in the dashed box form a single layer, which can be repeated several times to enhance expressibility. The parameters in different layers are independent.} 
    \label{fig:ansatz}
\end{figure}

In this work, we consider a problem-independent, hardware-efficient ansatz denoted by a joint unitary gate $U(\boldsymbol{\theta})$. In Fig.~\ref{fig:ansatz}, we provide a concrete implementation of $U(\boldsymbol{\theta})$ applied in our numerical simulation, which is composed of $R_y$, $R_z$, and controlled-$Z$ gates. We assume each single-qubit gate is
parameterized by a single scalar $\theta_i$, representing the $i$-th component of $\boldsymbol{\theta}$. The gates in the dashed box form a single layer, which
can be repeated several times to enhance expressibility. The parameters in different layers are independent.

Now, we discuss how to estimate complex-valued and real-valued quadratic forms by performing measurements on the quantum circuit. For the complex-valued case, we consider encoding the complex vector $\boldsymbol{x}$ in the amplitude of the parameterized quantum state $\ket{\psi(\boldsymbol{\theta})}$. Since the quantum state should satisfy the normalization constraint $\braket{\psi(\boldsymbol{\theta})}{\psi(\boldsymbol{\theta})}=1$, we introduce an additional variable $\eta \in \mathbb{R}$ as the normalization parameter, which leads to a normalized complex variable $\|\eta^{-1/2}\boldsymbol{x}\|_2 = 1$. Then, the parameterized quantum state $\ket{\psi(\boldsymbol{\theta})}$ can be expressed as
\begin{equation}\label{eq:amplitude_encoding}
\ket{\psi(\boldsymbol{\theta})} = \eta^{-\frac{1}{2}}\sum_{j=1}^N x_j \ket{\phi_j}, 
\end{equation}
where $\{\ket{\phi_j}\}_{j=1}^{N}$ is the computational basis. 
Then it is straightforward that the expectation of an observable $A_i$, $\bra{\psi(\boldsymbol{\theta})}A_i\ket{\psi(\boldsymbol{\theta})}$ is a quadratic form of the complex vector $\boldsymbol{x}$
\begin{equation}\label{eq:complex_quadratic_form}
\boldsymbol{x}^\dag A_i \boldsymbol{x}= \eta \bra{\psi(\boldsymbol{\theta})}A_i\ket{\psi(\boldsymbol{\theta})}.
\end{equation}
In the following, we use the notation $\boldsymbol{\vartheta} = (\eta,\boldsymbol{\theta})$ as a concatenation of $\eta$ and $\boldsymbol{\theta}$.

For the real-valued case, one needs to measure $\ket{\psi(\boldsymbol{\theta})}$ in the computational basis to obtain a real-valued diagonal density matrix $\rho_{\mathrm{diag}}$, whose $j$-th diagonal element is $\rho_{\mathrm{diag},j} =|\bra{\phi_j}\ket{\psi(\boldsymbol{\theta})}|^2$. Then the real vector $\boldsymbol{y}$ is encoded in the probabilities of measurement outcomes with a normalization parameter $\eta$ such that $\eta^{-1/2}\sum_{j=1}^N y_j = 1$.
Then
\begin{equation}\label{eq:real_quadratic_form}
\begin{aligned}
\boldsymbol{y}^\dag B_i \boldsymbol{y} & = \sum_{k,l=1}^N B_{i,kl}y_k y_l\\
& = \eta \sum_{k,l=1}^N B_{i,kl}|\bra{\phi_k}\ket{\psi(\boldsymbol{\theta})}|^2 |\bra{\phi_l}\ket{\psi(\boldsymbol{\theta})}|^2.
\end{aligned}
\end{equation}
To estimate the quadratic form, we apply the following Monte Carlo method in Algorithm~\ref{alg:monte-carlo}.


\begin{algorithm}[htb]\caption{Monte-Carlo Method for Estimating a Real-Valued Quadratic Form}\label{alg:monte-carlo}
\KwIn{circuit parameter $\boldsymbol{\theta}$, number of samples $S$.}
\textbf{For} $i=1:1:S$ \textbf{do}  \\
\quad Run the quantum circuit to prepare $\ket{\psi(\boldsymbol{\theta})}$; \\

\quad Measure $\ket{\psi(\boldsymbol{\theta})}$ in the computational basis and record the measurement output $k$; \\

\quad Run the quantum circuit to prepare $\ket{\psi(\boldsymbol{\theta})}$; \\

\quad Measure $\ket{\psi(\boldsymbol{\theta})}$ in the computational basis and record the measurement output $l$; \\
\textbf{end For} \\
Calculate the estimator $\eta \cdot \sum_{i=1}^S B_{i,kl}/S$.
\end{algorithm}

Due to the non-negativity of a density matrix, Eq.~\eqref{eq:real_quadratic_form} represents a quadratic form with non-negative variables. Besides, note that any general real-valued QCQP can be transformed into a QCQP with non-negative variables by simply replacing each variable $y \in \mathbb{R}^N$ with the difference of two non-negative variables $y_1 - y_2$, where $y_1, y_2 \geq 0$.

And
\begin{equation}
\boldsymbol{y}^\dag B_i \boldsymbol{y} =
\left[\begin{array}{cc}
\boldsymbol{y_1}^\dag & \boldsymbol{y_2}^\dag  
\end{array}
\right]
\left[ 
\begin{array}{cc}
B_i & - B_i \\
-B_i & B_i  \\
\end{array}
\right]
\left[\begin{array}{c}
\boldsymbol{y_1} \\
\boldsymbol{y_2}  \\
\end{array}
\right].
\end{equation}
Therefore, our method will not lose generality.

According to Eq.~\eqref{eq:amplitude_encoding}, we can see that the number of qubits has a logarithmic scaling in terms of the original variable size $N$. On the other hand, the parameters in a hardware-efficient ansatz scale polynomially with the number of qubits. Then we have $n \sim \mathrm{polylog}(N)$, which means the variational algorithm greatly simplifies the computation cost when $N$ is large enough. This fact also holds for the real-valued case.

\subsection{Hybrid Algorithm}

In this section, we present the hybrid algorithm.
We unify Eq.~\eqref{eq:qcqp} and Eq.~\eqref{eq:qcqpreal} into the following form,
\begin{equation}\label{eq:formulationre}
\begin{aligned}
\mathrm{min}_{\boldsymbol{\vartheta}} \quad &  F_0(\boldsymbol{\vartheta}) \\
\mathrm{s.t.} \quad & F_i(\boldsymbol{\vartheta})\leq 0,  \quad i\in\{1,2,\dotsc, m\}  \\
\end{aligned}
\end{equation}
where $F_i: \mathbb{R}^{n+1} \rightarrow \mathbb{R}, \forall i \in\{0,1,\dotsc, m\}$. The explicit form of $F_i(\boldsymbol{\vartheta})$ depends on a complex-valued or real-valued QCQP according to Eqs.~\eqref{eq:complex_quadratic_form} and \eqref{eq:real_quadratic_form}, respectively.
The Lagrange function is,
\begin{equation}\label{eq:lagrange}
 L(\boldsymbol{\vartheta},\boldsymbol{\lambda})=   F_0(\boldsymbol{\vartheta}) + \sum_{i=1}^m \lambda_i F_i (\boldsymbol{\vartheta}),
\end{equation}
where $\boldsymbol{\lambda} = (\lambda_1,\lambda_2, \cdots, \lambda_m)$ is called dual variable.
Then the minimization problem
Eq.~\eqref{eq:formulationre} can be expressed as the following min-max problem
\begin{equation}\label{eq:maxminl}
\min_{\boldsymbol{\vartheta}\in \mathbb{R}^{n+1}} \max_{\boldsymbol{\lambda} \in\mathbb{R}_{\geq 0}^{m}}L( \boldsymbol{\vartheta},\boldsymbol{\lambda}).
\end{equation}
In the classical part of the variational quantum algorithm, we use a primal-dual interior point method for the non-convex optimization \cite{Wchter2006OnTI} to optimize primal variables $\boldsymbol{\vartheta}$ and dual variables $\boldsymbol{\lambda}$ simultaneously. This algorithm is based on second-order gradients of the quadratic forms. Thanks to the existence of parameter-shift rule in hardware-efficient ansatz \cite{PhysRevA.98.032309,PhysRevA.99.032331,zhou2023hybrid}, the first and second order gradients can be estimated efficiently.

The basic idea of the primal-dual interior point method is to solve the perturbed Karush-Kuhn-Tucker (KKT) condition of the Lagrange function Eq.~\eqref{eq:lagrange}, where the complementary slackness is modified to involve a parameter $\mu$, $\lambda_i F_i(\boldsymbol{\vartheta})= -\mu, \forall i$. Such a modification is to make the KKT condition smooth and easier to solve. In the algorithm, the perturbation parameter $\mu$ is contracted to zero from an initial value $\mu_0$ iteratively, which can be intuitively understood as a navigation of the search direction.
When $\mu \rightarrow 0$, the solution will converge to that of the original problem without perturbation.


When $\mu$ is given,
the key step is to calculate the search direction by solving the KKT condition, which is given by a linear Newton system
\begin{equation}\label{eq:search}
\left[\begin{array}{cc}
\nabla^2 F_0(\boldsymbol{\vartheta}) + \sum_{i=1}^m \lambda_i \nabla^2 F_i(\boldsymbol{\vartheta})& \nabla \textbf{F} (\boldsymbol{\vartheta})^T   \\
-\mathrm{diag}(\boldsymbol{\lambda}) \nabla \textbf{F} (\boldsymbol{\vartheta}) &-\mathrm{diag}(\textbf{F} (\boldsymbol{\vartheta}))  \\

\end{array}
\right]
\left[ 
\begin{array}{c}
\Delta \boldsymbol{\vartheta} \\
\Delta \boldsymbol{\lambda} \\
\end{array}
\right]
= \boldsymbol{r}.
\end{equation} 
where 
$\boldsymbol{r}$ is the vector of residuals
\begin{equation}
\boldsymbol{r} = \left[ 
\begin{array}{c}
\nabla F_0(\boldsymbol{\vartheta}) + \sum_{i=1}^m \lambda_i \nabla F_i(\boldsymbol{\vartheta})  \\
-\mathrm{diag}(\boldsymbol{\lambda}) \textbf{F} (\boldsymbol{\vartheta}) - \mu \boldsymbol{e}
\end{array}
\right] =
\left[
\begin{array}{c}
r_{\mathrm{dual}} \\
r_{\mathrm{cent}} \\
\end{array}
\right]
.
\end{equation}
We explain the notations in the linear equations above. The column vector $\textbf{F} (\boldsymbol{\vartheta}) =  [F_1(\boldsymbol{\vartheta}), F_2(\boldsymbol{\vartheta}), \dotsc, F_m(\boldsymbol{\vartheta})]^T$; The diagonal matrix $\mathrm{diag}(\boldsymbol{\lambda}) = \sum_{i=1}^m \lambda_i \ket{i}\bra{i}$; $\mathrm{diag}(\textbf{F} (\boldsymbol{\vartheta})) = \sum_{i=1}^m F_i (\boldsymbol{\vartheta}) \ket{i}\bra{i}$; The column vector $\boldsymbol{e} = (1,1,\dotsc, 1)^T$. Then $\nabla \textbf{F} (\boldsymbol{\vartheta})$ is a Jacobian matrix.

Then the primal and dual variables are updated by iterations. Suppose the current iteration step is labeled as $k$.
After determining the search directions, we also need to find an appropriate iteration step length $\alpha^{(k)}$ by line search \cite{Boyd2006ConvexO,Wchter2006OnTI}.
Then the parameters can be updated by the following rules,
\begin{equation}\label{eq:update}
\begin{aligned}
\boldsymbol{\vartheta}^{(k+1)} = \boldsymbol{\vartheta}^{(k)} + \alpha^{(k)} \Delta \boldsymbol{\vartheta}^{(k)}  \\
\boldsymbol{\lambda}^{(k+1)} = \boldsymbol{\lambda}^{(k)} + \alpha^{(k)} \Delta\boldsymbol{\lambda}^{(k)}  \\
\end{aligned}
\end{equation}
The algorithm will terminate when the residuals are small enough. Define a quantity
\begin{equation}
R_\mu (\boldsymbol{\vartheta},\boldsymbol{\lambda}) = \max\left(||r_{\mathrm{dual}}||_\infty,||r_{\mathrm{cent}}||_\infty \right).
\end{equation}
The termination criteria for the iterations with each $\mu$ is given by 
\begin{equation}\label{eq:termination1}
R_\mu (\boldsymbol{\vartheta},\boldsymbol{\lambda}) < c_\epsilon \mu,
\end{equation}
for some given constant $c_\epsilon$. When Eq.~\eqref{eq:termination1} is satisfied, $\mu$ is updated by multiplying by a constant $c_\mu \in (0,1)$ and eventually goes to $0$.  
The termination criteria for the whole algorithm is 
\begin{equation}\label{eq:termination2}
R_0 (\boldsymbol{\vartheta},\boldsymbol{\lambda}) < \epsilon,
\end{equation}
for some given constant $\epsilon$. The hybrid algorithm is summarized in Algorithm~\ref{alg:interior}. Note that 
the variable $\eta$ in $\boldsymbol{\vartheta}$ is not a circuit parameter. The corresponding partial derivative is calculated numerically rather than using parameter-shift rule.
\begin{algorithm}[htb] \caption{Hybrid algorithm for QCQP} \label{alg:interior}
\KwIn{initial point $(\boldsymbol{\vartheta}^{(0)}, \boldsymbol{\lambda}^{(0)})$, parameters for termination criteria $\epsilon$ and $c_\epsilon$, initial barrier parameter $\mu_0$ and its scaling parameter $c_\mu$.}
	
Initialize iteration counter $l\leftarrow 0$ for outer loops and $k\leftarrow 0$ for inner loops. \\
\textbf{While} $R_{\mu^{(l)}}(\boldsymbol{\vartheta}^{(k)},\boldsymbol{\lambda}^{(k)}) \geq \epsilon$ \textbf{do}  \\
 \quad \textbf{While} $R_{\mu^{(l)}}(\boldsymbol{\vartheta}^{(k)},\boldsymbol{\lambda}^{(k)}) \geq c_\epsilon \mu$ \textbf{do} \\

\quad\quad Calculate the following quantities by the quantum computer, $F_i(\boldsymbol{\vartheta}^{(k)})$ $(i\in \{0,1,\dotsc, m\})$, $\nabla F_0 (\boldsymbol{\vartheta}^{(k)})$, $\nabla^2 F_0 (\boldsymbol{\vartheta}^{(k)})$, $\nabla \textbf{F}(\boldsymbol{\vartheta}^{(k)})$ and $\nabla^2 \textbf{F}(\boldsymbol{\vartheta}^{(k)})$. \\
\quad\quad Compute search direction by Eq.~\eqref{eq:search}. \\
\quad\quad Determine the step length $\alpha^{(k)}$ by line search. \\ 
\quad\quad Update primal and dual variables by Eq.~\eqref{eq:update}. \\
\quad \textbf{end While} \\
\quad Update the perturbation parameter by $\mu^{(l+1)} \leftarrow c_\mu \mu^{(l)}$ \\
\textbf{end While}
\end{algorithm}


\section{Examples}

\subsection{Optimal Power Flow Problem}
\begin{figure*}[htbp]
    \centering
    \includegraphics[width=0.75\textwidth]{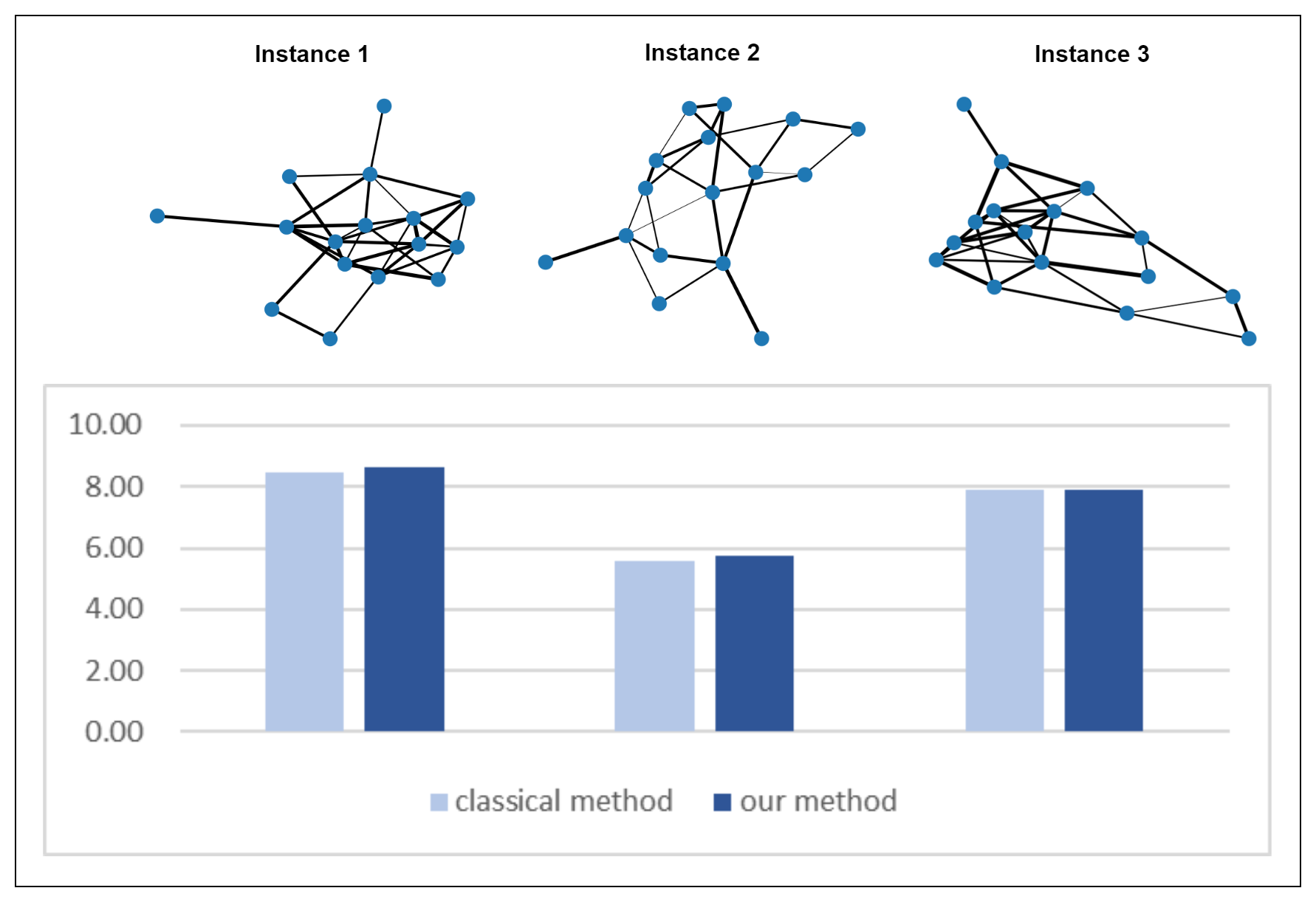}
    \caption{Numerical experiment results of our algorithm for solving random OPF problem instances. The graphs are randomly generated, where the width of the edges reflects the admittance. We list the results solved by our algorithm and a classical non-linear optimizer in the histogram. Compared to the classical algorithm, our algorithm performs better in these random instances.}
    \label{fig:opf-result}
\end{figure*}

The Optimal Power Flow (OPF) problem is a fundamental optimization challenge in electrical power systems engineering. It involves determining the optimal operating conditions for a power system, such as the generation and distribution of electrical power, while satisfying various engineering constraints and objectives. The primary goal of OPF is to minimize the cost of electricity generation while ensuring the system operates within acceptable limits to maintain reliability and stability.
Mathematically, the OPF problem can be formulated as a complex-valued QCQP problem \cite{Frank2016AnIT},
\begin{equation}
\begin{aligned}
\min_{\boldsymbol{x}\in\mathbb{C}^{N}} \quad & \boldsymbol{x}^\dagger Y' \boldsymbol{x} + \sum_{j=1}^{n} \operatorname{Re} S_j^L
\\
\mathrm{s.t.} \quad & v^\dagger Y'_j \boldsymbol{x} + \operatorname{Re} S_j^L\in [0,1], \quad j\in\{1,2,\dotsc,N\}
\\
     \quad & \boldsymbol{x}^\dagger Y''_j \boldsymbol{x}+ \operatorname{Im} S_j^L\in [0,1], \quad j\in\{1,2,\dotsc,N\}
\\
     \quad & \boldsymbol{x}^\dagger E_{jj} \boldsymbol{x} \in [0,1], \quad j\in\{1,2,\dotsc,N\}.
\end{aligned}
\end{equation}
Here, $S$ is a vector representing the complex power flow, $Y'$ and $Y''$ are relevant to the admittance matrix of the electrical system. They are all constants determined by the input electrical power system. The binary matrix $E_{kl}$ has only one non-zero entry at the $k$-th row and $l$-th column.

Our numerical experiment is shown in Fig.~\ref{fig:opf-result}. The graphs are randomly generated, representing the topological structure of electrical systems. The width of an edge reflects the value of admittance of the transmission line. We ran the OPF problem using our hybrid algorithm and a classical optimizer widely applied in non-linear optimizations, named Interior Point OPTimizer (IPOPT) \cite{IPOPT}. We set the number of layers in the hardware-efficient ansatz in Fig.~\ref{fig:ansatz} to be 5. The hybrid algorithm outperforms the classical optimizer in these instances.
We refer to Appendix \ref{app:experiment} for details on the problem formulation and the generation of random instances.

\subsection{Max-Cut Problem}
The Maximum Cut (Max-Cut) problem is a well-known combinatorial optimization problem with applications in various fields, including computer science, graph theory, and network design. In the Max-Cut problem, given an undirected graph, the goal is to partition its vertices into two sets, such that the number of edges crossing between the two sets is maximized. In other words, we want to find the cut that separates the graph into two disjoint sets with the maximum number of edges crossing the cut.
Mathematically, the Max-Cut problem can be formulated as a real-valued QCQP problem,
\begin{equation}
\begin{aligned}
    \min_{\boldsymbol{y}\in\mathbb{R}^N} \quad & -\frac{1}{2}(M-\boldsymbol{y}^\dagger A \boldsymbol{y}) \\
    \mathrm{s.t.} \quad & \boldsymbol{y}^\dagger E_{jj} \boldsymbol{y} = 1, \quad j\in\{1,2,\dotsc,N\},
\end{aligned}
\end{equation}
where $A$ is the adjacency matrix of the input graph $G(V,E)$, and $M=|E|$ is the total number of edges,
\begin{align}
    A_{ij}=\begin{cases}
        1, \quad (i,j)\in E \\
        0, \quad \text{else}
    \end{cases}
\end{align}

\begin{figure*}[htbp]
    \centering
    \includegraphics[width=0.75\textwidth]{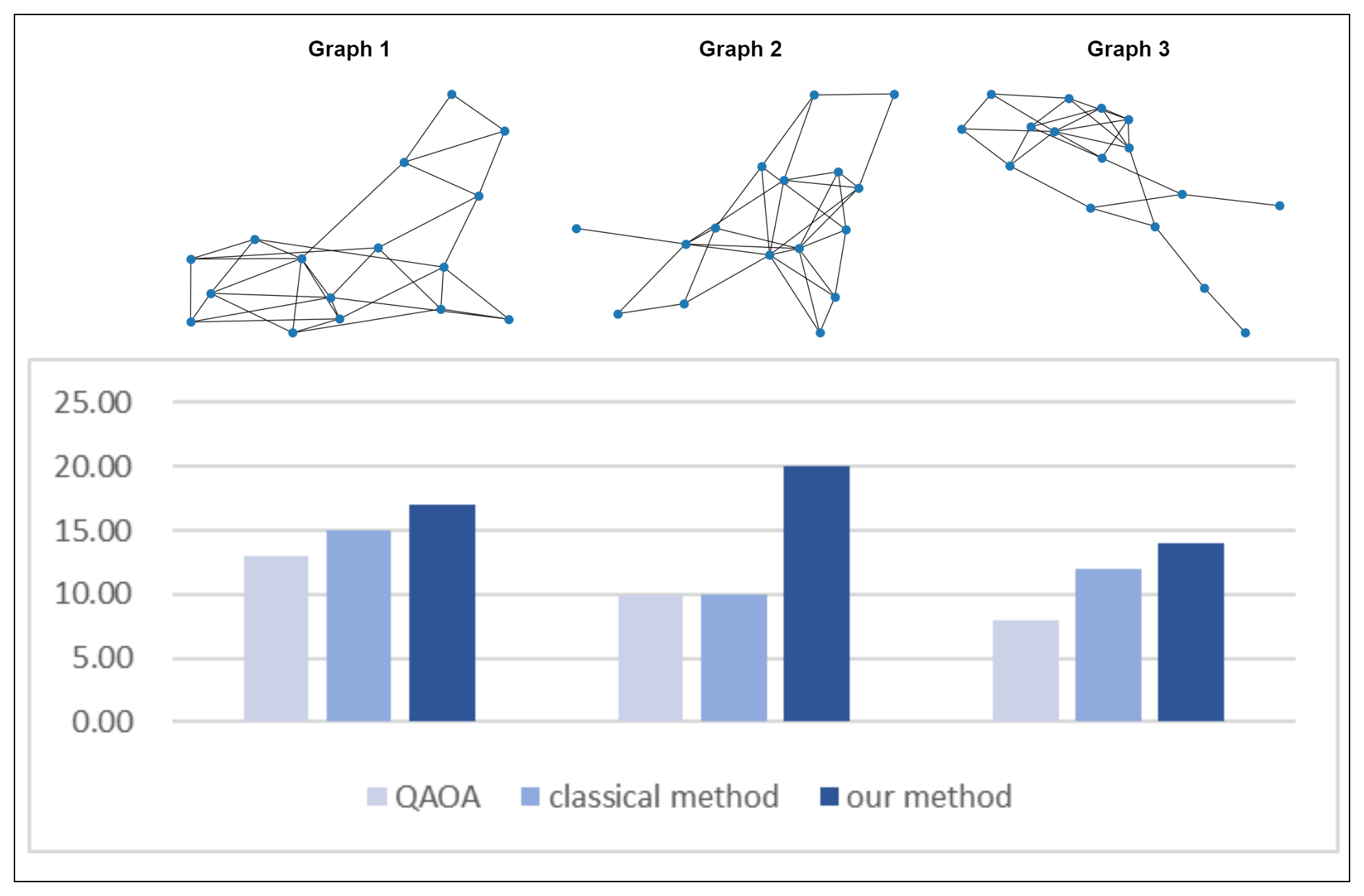}
    \caption{Numerical experiment results of our algorithm on solving the Max-Cut problem on randomly generated graphs. We compare our results with those calculated by QAOA and a classical non-linear optimizer. It turns out that our algorithm outperforms the other two algorithms on these random graphs.}
    \label{fig:max-cut-result}
\end{figure*}

Our numerical experiment is shown in Fig.~\ref{fig:max-cut-result}, where the graphs are also randomly generated. The Max-Cut problem is a quadratic unconstrained binary optimization (QUBO) problem, which can be solved by QAOA. Therefore, besides the classical optimizer, we also compare our results with those obtained by QAOA. The number of layers $p$ in QAOA is chosen to be 4, surpassing the depth of the hardware-efficient ansatz employed in our algorithm. It turns out that our algorithm outperforms the other two algorithms in these random graphs.
We refer to Appendix \ref{app:experiment} for details on the generation of random graphs.

\section{Conclusion}
In conclusion, we have proposed a variational quantum algorithm for QCQPs which can outperform its classical counterpart. Compared to existing quantum algorithms, our algorithm can handle QCQPs with general quadratic constraints and significantly reduce the quantum resource requirement. For future work, one may apply variational quantum algorithms (VQAs) to handle more general non-linear, non-convex optimization problems beyond QCQPs. Moreover,
the exploration of more efficient classical optimization algorithms, such as quasi-Newton methods, could accelerate the optimization process in VQAs.
Finally, optimization theory can be leveraged to refine the classical part of VQAs. Integrating optimization principles into the design and implementation of VQAs has the potential to yield more robust and efficient algorithms.

\section*{Acknowledgement}
We thank Guojing Tian for insightful discussions. This work was supported in part by the National Natural Science Foundation of China Grants No. 61832003, 62272441, 12204489, and the Strategic Priority Research Program of Chinese Academy of Sciences Grant No. XDB28000000. Qian Li's work was additionally supported by Hetao Shenzhen-Hong Kong Science and Technology  Innovation Cooperation Zone Project (No.HZQSWS-KCCYB-2022046).

\onecolumngrid
\appendix

\section{Primal-dual interior-point method}
In this section, we briefly introduce the primal-dual interior-point method. 
Based on the Lagrange function in Eq.~\eqref{eq:lagrange}, we first calculate the original Karush-Kuhn-Tucke (KKT) condition. The KKT conditions are a set of necessary conditions that any solution to a constrained optimization problem must satisfy in order to be optimal. This means that if a point satisfies the KKT conditions, it has a strong indication of being a candidate for the optimal solution. The KKT condition of Eq.~\eqref{eq:formulationre} is given by
\begin{equation}\label{eq:KKTcondition}
\begin{aligned}
\nabla F_0(\boldsymbol{\vartheta}) + \sum_{i=1}^m  \lambda_i \nabla F_i (\boldsymbol{\vartheta}) & = 0 \\
F_i(\boldsymbol{\vartheta}) & \leq 0 \\
\lambda_i  &\geq 0 \\
\lambda_i F_i(\boldsymbol{\vartheta}) & = 0.
\end{aligned}
\end{equation}
The first line means the gradient should vanish for the optimal solution. The second, third and last lines stand for the primal feasibility, dual feasibility and complementary slackness, respectively.   

The central idea of the primal-dual interior-point method is to solve a perturbed KKT condition where the complementary slackness slighted modified to
\begin{equation}
\lambda_i F_i(\boldsymbol{\vartheta})  = -\mu.
\end{equation}
The parameter $\mu$ has a stabilizing effect on the optimization process. As $\mu$ decreases, the optimization algorithm approaches a solution that satisfies the original KKT conditions, effectively moving from the infeasible region toward the feasible region. This controlled movement ensures that the algorithm remains stable and converges efficiently.
Then we solve the perturbed KKT condition by solving the linear system,
\begin{equation}
\begin{aligned}
\boldsymbol{r}(\boldsymbol{\vartheta}, \boldsymbol{\lambda}) = \left[ 
\begin{array}{c}
\nabla F_0(\boldsymbol{\vartheta}) + \sum_{i=1}^m \lambda_i \nabla F_i(\boldsymbol{\vartheta})  \\
-\mathrm{diag}(\boldsymbol{\lambda}) \textbf{F} (\boldsymbol{\vartheta}) - \mu \boldsymbol{e}
\end{array}
\right] =
0.
\end{aligned}
\end{equation}
We use the Newton's method to calculate the search direction $(\Delta \boldsymbol{\vartheta}, \Delta \boldsymbol{\lambda})$,
\begin{equation}\label{eq:Newton1}
\begin{aligned}
\boldsymbol{r}(\theta + \Delta \boldsymbol{\vartheta}, \Delta \boldsymbol{\lambda}) =\boldsymbol{r}(\boldsymbol{\vartheta}, \boldsymbol{\lambda}) + D\boldsymbol{r}  \left[ 
\begin{array}{c}
 \boldsymbol{\vartheta}  \\
\boldsymbol{\lambda}
\end{array} \right] =0,
\end{aligned}
\end{equation}
where 
\begin{equation}\label{eq:Newton2}
\begin{aligned}
D\boldsymbol{r} = \left[\begin{array}{cc}
\nabla^2 F_0(\boldsymbol{\vartheta}) + \sum_{i=1}^m \lambda_i \nabla^2 F_i(\boldsymbol{\vartheta})& \nabla \textbf{F} (\boldsymbol{\vartheta})^T   \\
-\mathrm{diag}(\boldsymbol{\lambda}) \nabla \textbf{F} (\boldsymbol{\vartheta}) &-\mathrm{diag}(\textbf{F} (\boldsymbol{\vartheta}))  \\
\end{array}
\right].
\end{aligned}
\end{equation}
Combining Eqs.~\eqref{eq:Newton1} and \eqref{eq:Newton2}, we can obtain Eq.~\eqref{eq:search} in the main text.
\section{Numerical experiment}\label{app:experiment}
In this section we present the details in the numerical experiment.
For the max-cut problem, we consider input graphs sampled from $G_{N,P}$, i.e., there are $N$ nodes in the graph and the probability of the existence of each edge is $P$. In our numerical experiment of max-cut problem, we choose $N=16$ and $P=0.25$.


For the OPF problem, the input electrical power system can be formulated as follows. Consider a graph $G$, where nodes represent buses and edges represent transmission lines. Each node is equipped with an admittance $y_k$ and the edge between node $j$ and node $k$ is equipped with an admittance $y_{jk}$. The admittance matrix $Y\in\mathbb{C}^{n\times n}$ is denoted as
\begin{equation}
\begin{aligned}
Y_{kk} &= y_k+\sum_{j\in N(k)}y_{jk}, \quad & k \in\{1,2, \ldots, N\} \\
Y_{jk} &= Y_{kj} = -y_{jk}, \quad  & j,k \in\{1,2, \ldots, N \},j\neq k.
\end{aligned}
\end{equation}
For the power flow at node $i$, the power flow $S_i^G$ generated at that node is equal to the power flow $S_i^L$ consumed at that node plus the output power flow, 
\begin{align}
S_i^G-S_i^L = \sum_{k=1,\ldots,N}\boldsymbol{x}_k^\dagger Y_{jk}\boldsymbol{x}_j,
\end{align}
where the voltage $\boldsymbol{x}\in \mathbb{C}^{N}$ stands for the complex voltage. Then we can formulate the original OPF problem:
\begin{equation}
\begin{aligned}
&\min_{\boldsymbol{x}\in\mathbb{C}^{N},S^G\in\mathbb{C}^{N}} & \sum_i \operatorname{Re}S_i^G &
\\
&\mathrm{s.t.} & S_j^G-S_j^L -\sum_{k=1\cdots N} \boldsymbol{x}_j Y_{jk}^\dagger \boldsymbol{x}_k^\dagger =0,
& \quad j\in\{1,2,\dotsc, N\}
\\
&& \operatorname{Re}S_j^G\in [\operatorname{Re}S_j^{G,\min},\operatorname{Re}S_j^{G,\max}],
&\quad j\in\{1,2,\dotsc, N\}
\\
&& \operatorname{Im}S_j^G\in [\operatorname{Im}S_j^{G,\min},\operatorname{Im}S_j^{G,\max}],
&\quad j\in\{1,2,\dotsc, N\}
\\
&& \|\boldsymbol{x}_j\| \in [\boldsymbol{x}_j^{\min},\boldsymbol{x}_j^{\max}],
& \quad j\in\{1,2,\dotsc, N\}.
\end{aligned}
\end{equation}
To demonstrate our algorithm, the formulation above is simplified. We set the range of $\operatorname{Re}S^G_j$ and
$\operatorname{Im}S^G_j$ as $[0,1]$ and representing $S^G$ in terms of $\boldsymbol{x}$:
\begin{equation}
\begin{aligned}
& \min_{ \boldsymbol{x}\in\mathbb{C}^N} & \boldsymbol{x}^\dagger Y' \boldsymbol{x} + \sum_j \operatorname{Re} S_j^L
\\
& \mathrm{s.t.} & \boldsymbol{x}^\dagger Y'_j \boldsymbol{x} + \operatorname{Re} S_j^L\in [0,1]
\\
&      & \boldsymbol{x}^\dagger Y''_j \boldsymbol{x}+ \operatorname{Im} S_j^L\in [0,1]
\\
&      & \boldsymbol{x}_j^\dagger \boldsymbol{x}_j \in [0,1],
\end{aligned}
\end{equation}
where $Y^\prime$ and $Y^{\prime\prime}$ are real and imaginary part of the admittance matrix $Y$, respectively. And
\begin{equation}
\begin{aligned}
Y'_j &= \begin{bmatrix}
    0 & \cdots & Y'_{j1} & \cdots & 0 \\
    0 & \cdots & Y'_{j2} & \cdots & 0 \\
    \cdots \\
    0 & \cdots & Y'_{jn} & \cdots & 0 \\
\end{bmatrix}\\
Y''_j &= \begin{bmatrix}
    0 & \cdots & Y''_{j1} & \cdots & 0 \\
    0 & \cdots & Y''_{j2} & \cdots & 0 \\
    \cdots \\
    0 & \cdots & Y''_{jn} & \cdots & 0 \\
\end{bmatrix}.
\end{aligned}
\end{equation}
In summary, the instances are randomly generated such that
\begin{enumerate}
    \item Each term of $\operatorname{Re} S^L,\operatorname{Im} S^L$ is a uniform random variable in the range of $[0,1]$;
    \item The graph $G$ is a randomly generated connected graph;
    \item The admittance $y_k,y_{jk}$ are also uniform random variables in the range of $[0,1]$, and then the corresponding admittance matrix can be generated.
\end{enumerate}

In the numerical experiment of both problems, the classical method refers to the widely applied interior point optimizer for non-linear optimizations (IPOPT) \cite{IPOPT}. While in our hybrid algorithm, the gradient, Hessian, and Jacobian matrix are obtained from the quantum processor. The optimization process is also accomplished by the primal-dual interior-point method in IPOPT. For the max-cut problem, we also compare our results with the well-known QAOA algorithm, using implementation from Qiskit library \cite{solving-qaoa}.
We make a comparison of the circuit depth between our algorithm and QAOA. The depth of a $l$-layer hardware-efficient ansatz used in our method is $4l+2$. While the depth of a $p$-layer QAOA ansatz is $p(d+1)$ where $d$ refers to maximal degree of one vertex in the graph. In our examples $l=5$ and $p=4$, and the example graph has maximal degree $d = 5$. Therefore, our algorithm can achieve a better performance in solving the Max-Cut problem for the random graphs with much fewer qubits and lower depth.
\bibliographystyle{apsrev4-1}

\bibliography{QCQPref}
\end{document}